\documentclass[twocolumn,nopacs,preprintnumbers,amsmath,amssymb]{revtex4}

\usepackage{verbatim}
\usepackage{graphicx}
\usepackage{dcolumn}
\usepackage{bm}

\begin{document}

\title{Charge Transfer Induced Polarity Switching in Carbon Nanotube Transistors}
\author{Christian Klinke}
\email{klinke@chemie.uni-hamburg.de}
\author{Jia Chen}
\author{Ali Afzali}
\author{Phaedon Avouris}
\affiliation{IBM Research Division, T. J. Watson Research Center, PO Box 218, Yorktown Heights, New York 10598}

\begin{abstract}

We probed the charge transfer interaction between the amine-containing molecules: hydrazine, polyaniline and aminobutyl phosphonic acid, and carbon nanotube field effect transistors (CNTFETs). We successfully converted p-type CNTFETs to n-type and drastically improved the device performance in both the ON- and OFF- transistor states utilizing hydrazine as dopant. We effectively switched the transistor polarity between p- and n- type by accessing different oxidation states of polyaniline. We also demonstrated the flexibility of modulating the threshold voltage (V$_{th}$) of a CNTFET by engineering various charge-accepting and -donating groups in the same molecule.

\end{abstract}

\maketitle

Carbon nanotube (CNT) based transistors have seen significant advances recently in terms of both understanding their interaction with the environment and their performance limits$^{1-3}$. Unlike conventional transistors made from bulk materials, the single atomic layer thin channel of a carbon nanotube field effect transistor (CNTFET) leads to an extreme sensitivity to its environment$^{4-6}$. In particular, charge transfer interaction between adsorbed atoms or molecules and the CNTFETs can modify their electronic characteristics$^{4-15}$. For example, electron donating effects to CNTs have been observed in amine-containing molecules like ammonia, butylamine, 3-(aminopropyl)-triethoxysilane$^{5}$, polyethyleneimine (PEI)$^{6}$, and proteins$^{7}$. These previous studies focused on the effect of electron transfer from the amine-containing molecules to CNTs on the ``ON''state of nanotube transistors with thick (unscaled) gate dielectrics, and attributed the device characteristic change to carrier density modification in the bulk of the CNTs. In this work, we study the effects of charge transfer from molecules to CNTFETs with scaled (thin) gate dielectrics on both the ON- and OFF-transistor states. The improvement in both the device drive current and the subthreshold behavior indicates a metal-CNT contact barrier reduction, not just charge transfer to the channel of the transistor. We also manipulate the charge transfer process by accessing different oxidation states of polyaniline and effectively switch the transistor polarity between p- and n- type. Furthermore, we demonstrate the flexibility of modulating the threshold voltage (V$_{th}$) of a CNTFET by engineering various charge-accepting and -donating groups in the same molecule: inducing negative V$_{th}$ change with electron accepting species and positive V$_{th}$ change after incorporating the electron donating amine group.

\begin{figure}[ht]
\begin{center}
\includegraphics[width=0.45\textwidth]{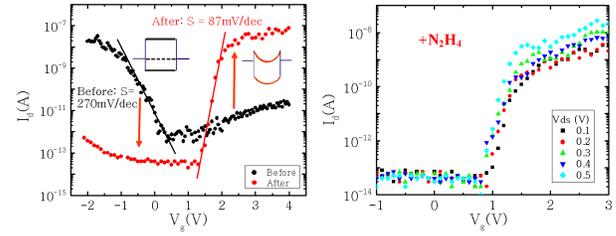}
\caption {\it a) Transfer characteristics of a CNTFET before and after n-doping at VV$_{ds}$ = 0.5~V with hydrazine.  The sub-threshold swing S improved from the original value of 270 to 87~mV/decade after doping. The inset shows a qualitative band diagram before and after doping. b) Transfer characteristics of a hydrazine-doped CNTFET at V$_{ds}$ = 0.1 to 0.5~V @ 0.1~V step.}
\end{center}
\end{figure}

We have fabricated CNTFETs using laser ablation CNTs, titanium source and drain electrodes separated by 300~nm on top of 10~nm SiO$^{2}$, and a Si backgate. Fabricated samples were then immersed into a 3~M solution of hydrazine (N$_{2}$H$_{4}$) in acetonitrile to form hydrazine-doped nanotube devices. The strong electron donating properties of hydrazine and its alkylated analogues are due to stabilizing effect of one amine group on the radical cation of the adjacent oxidized amine group$^{17}$. After doping, excess dopants were removed by rinsing with the solvent. Typical device transfer characteristics (I$_{ds}$ vs. V$_{gs}$) at V$_{ds}$ = 0.5~V of a CNTFET before and after doping are shown in Fig.~1. They show that: i) The p-type CNTFET was successfully converted to a n-type, confirming the electron transfer from hydrazine to the CNTFET; ii) The electron drive current Ion improved by 3 orders of magnitude and the contact resistance between the CNT and the metal was greatly reduced; iii) The minority carrier (hole current) injection was suppressed, and the OFF state at V$_{gs}$ = $+$1~V was improved by more than one order of magnitude after doping; iv) A significantly improved subthreshold swing S = dV$_{gs}$/d(logI$_{d}$) from the 270~mV/dec in the pristine device to the 87~mV/decade in the doped device, sharper compared with those reached either by K-doping$^{19}$, or by using Al as source/drain contacts$^{3}$; and v) The device V$_{th}$ has little dependence on the varying drain field as shown in Fig.~1b, unlike in an ambipolar CNTFET whose V$_{th}$ varies strongly with drain bias due to minority carrier injection$^{10}$. The much improved subthreshold slope of the hydrazine doped FET, together with the increase of electron current, suppression of hole current therefore improved OFF state, and independence of V$_{th}$ to the drain field, indicates a reduction of the electron injection SB between the CNT and the source/drain metal and an increase of the hole injection SB. The inset schematic band diagram of the device before and after doping in Fig.~1a shows the improved SB for electron injection, an increased SB for hole injection and some band bending due to transferred electrons from the dopants to the tube. The origin of the electron SB reduction could arise from interface dipole modification by the dopants at the metal-CNT contacts. After hydrazine donated electron to CNTs and forms N$_{2}$H$_{4}$$^{+}$, it induces negative image charge on the source/drain electrodes, forming inward-pointing surface dipole. The surface dipole reduces the local metal workfunction, which favors electron injection, suppresses hole injection and improves the OFF state of the device. The presence of dopants impacts both the carrier injection properties at the contacts and the carrier density in the bulk of the tube. We learned from previous work$^{20}$ that at a moderate doping density, modification on the contacts dominates; at high doping density, carrier density modification dominates. Most of previous work on charge transfer between amine-containing molecules and CNTs utilized large diameter tubes whose doping is more efficient, and charge transfer to the bulk of CNTs dominates. In this work, we work in the low doping regime where the interaction between dopants and contacts are dominant. In addition to the amine-containing monomer, we investigated amine-rich aliphatic polymers such as polyethyleneimine (PEI)$^{6}$ as n-dopants. Using these, we were able to tune the device V$_{th}$ by varying the doping concentration, and obtained enhancement as well as depletion mode transistors$^{20}$. We routinely stored the devices under nitrogen, and we observed reproducible doping results from them.

\begin{figure}[ht]
\begin{center}
\includegraphics[width=0.45\textwidth]{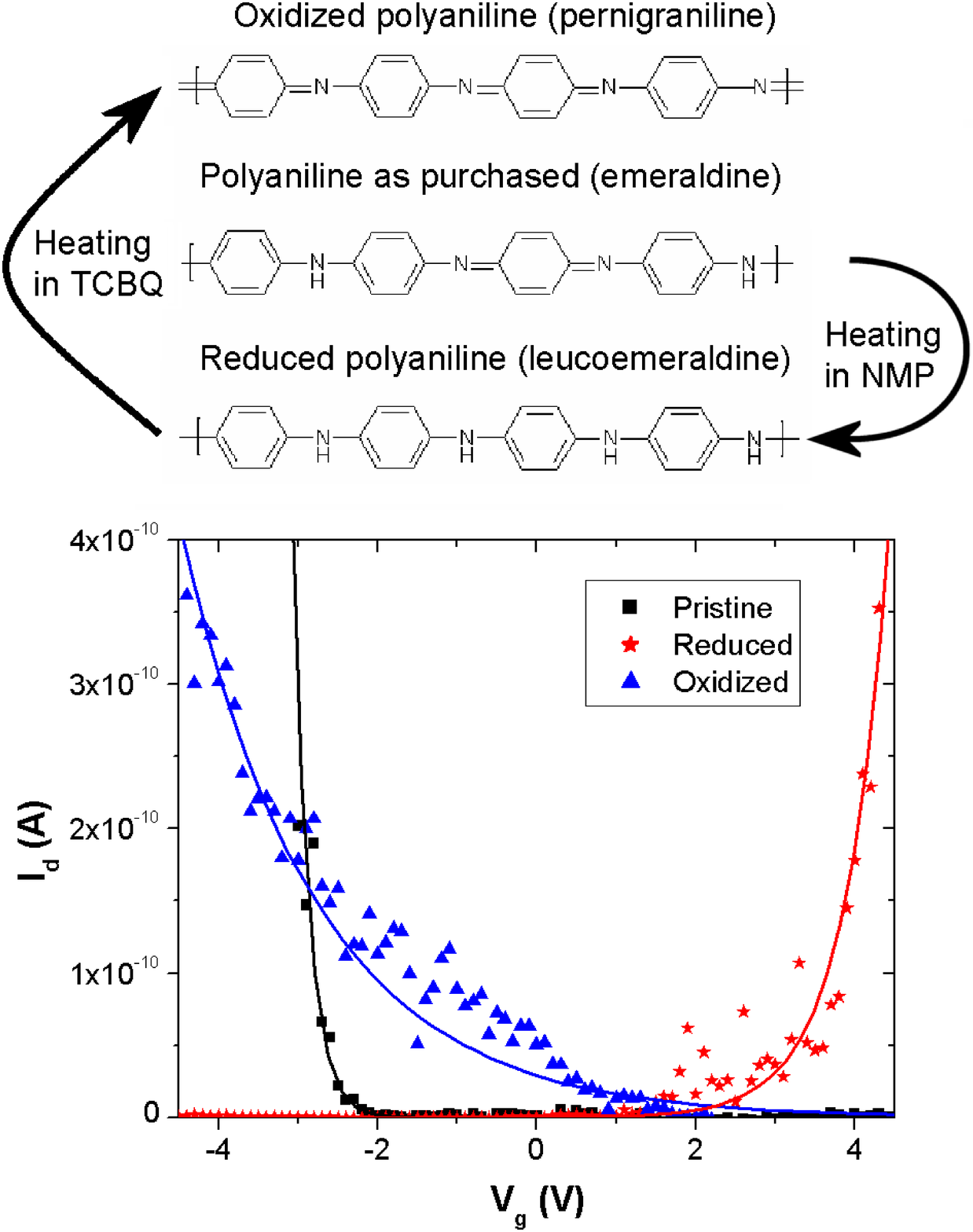}
\caption {\it a) Structure of the three oxidation states of polyaniline (PANI).  b) Transfer characteristics of a CNTFET after PANI (\textbf{L})-doping at VV$_{ds}$ = 0.5~V, which converted the CNTFET from p-type to n-type and back to p-type after PANI (\textbf{P}) doping of the same nanotube transistor.  The solid lines are meant as guides to the eyes.}
\end{center}
\end{figure}

We demonstrated above that the electron-donating amine-group is capable of transferring electrons to nanotubes and convert p-type to n-type CNTFETs. This conversion, however, cannot be reversed. In the following, we demonstrate the ability to switch between the p- and n-polarities of a CNTFET by controlling the charge transfer ability of the nitrogen using the different oxidation states of polyaniline. We employ the redox-active polyaniline (PANI) to modulate oxidation states of the polymer and characterize its interaction with CNTFETs. PANI has three distinct oxidation states: 1) the fully oxidized pernigraniline (\textbf{P}) has all its nitrogens in the form of imine groups with sp$^{2}$ hybridization, and thus has the lowest reduction potential (i.e., it is electron accepting); 2) The highly reduced leucoemeraldine (\textbf{L}) has all nitrogens in the form of a secondary amine with sp$^{3}$ hybridization, and has the lowest oxidation potential of the three states therefore a good reducing agent (i.e. it is electron donating); 3) the partially oxidized and partially reduced emeraldine (\textbf{E}) has a mixture of amine and imine groups and partially delocalized nitrogen 2p electrons (50\% sp$^{2}$ + 50\% sp$^{3}$ hybridization) (Fig.~2a). In the following we discuss transport measurements using these three states of PANI.

The CNTFETs were fabricated using again laser ablation CNTs, deposition of $\sim$1~nm Ti followed by 25~nm Pd to form the source and drain electrodes, a channel length of 500~nm, 20~nm SiO$_{2}$ as gate dielectric and a Si back-gate. The p-CNTFET transfer characteristic at V$_{ds}$ = $-$0.5~V is shown by the black curve in Fig.~2b. Then, we decorated the device using the fully reduced \textbf{L} with the more localized amine groups. In order to prepare \textbf{L}, we heated the as-purchased partially oxidized form \textbf{E} in N-methylpyrrolidinone (NMP) at 160$^{\circ}$C for 2~h in N$_{2}$ to fully reduce it$^{22}$. The solution was then spin-coated onto the CNTFET devices and heated at 160$^{\circ}$C in N$_{2}$ to drive out the solvent. The transfer characteristic of the CNTFET after \textbf{L} doping (red curve in Fig.~2b) shows the successful conversion of the original p-type to an n-type CNTFET. This p to n conversion after doping is consistent with the hydrazine and PEI doping results where the electron lone pair in the amine group donates electrons to CNTFETs and, in addition, modifies the metal-tube interface band line-up$^{20}$. To prevent device re-oxidation over time, we can spin-coat the sample with PMMA and bake it for an hour under N$_{2}$.

\begin{figure}[ht]
\begin{center}
\includegraphics[width=0.45\textwidth]{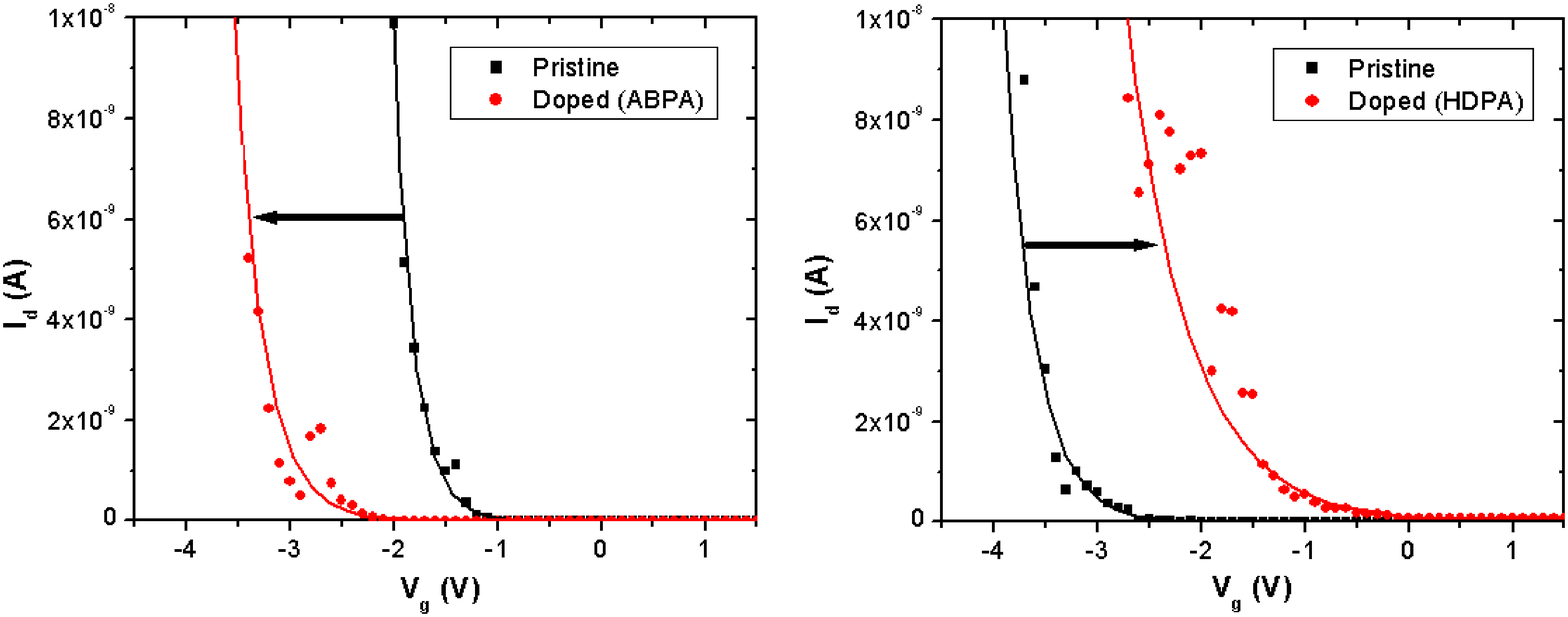}
\caption {\it a) Transfer characteristics of a CNTFET after aminobutyl phosphonic acid (ABPA) doping at V$_{ds}$ = 0.5~V, which shifts V$_{th}$ in the n-type direction. b) Transfer characteristics of a CNTFET after hexadecyl phosphonic acid (HDPA) doping at V$_{ds}$ = $-$0.5V, which shifts V$_{th}$ in the p-type direction.  The solid lines are meant as guides to the eyes.}
\end{center}
\end{figure}

In order to reduce the charge transfer from PANI to the CNTFET, we fully oxidized PANI to the state \textbf{P} where the nitrogen in the imine group forms a double bond with the quinoid ring and has sp$^{2}$ hybridization. The oxidation process was carried out by immersing the \textbf{L} covered device in a 1\% solution of tetrachloro-1,4-benzoquinone (TCBQ) in 4:1 acetonitrile/dimethylacetamide at 80$^{\circ}$C for 30~min. The transfer characteristic after the oxidation process is shown by the blue curve in Fig.~2b, where the CNTFET was successfully converted from n- back to p-type. The TCBQ molecules do not interact directly with the CNTFET because the devices were fully covered by PANI which does not dissolve in the solvent (4:1 acetonitrile/dimethylacetamide) where TCBQ was introduced. Unlike doping with hydrazine, the ON- and OFF- state currents of devices doped by the modified PANIs (\textbf{P} and \textbf{L}) are comparable to the undoped devices$^{23}$. Further optimization of device performance can be achieved by optimizing solvents (the NMP used in current work may have reduced the adhesion between the source/drain metal film and the CNT, therefore increasing the contact resistance), increasing doping density and reducing the gate dielectric thickness. The highly oxidized \textbf{P} acts as an electron-accepting molecule that p-dopes the nanotube transistor, and the V$_{th}$ is shifted to more positive values compared to the pristine device. Since \textbf{P} is the most stable form of PANI under ambient conditions, these p-doped devices are stable in air without protection. Therefore, by modifying the oxidation state of PANI from highly reduced \textbf{L} to completely oxidized \textbf{P}, we have demonstrated a polarity switching of a CNTFET from p- to n- then back to p-type. 

The correlation between the charge transfer ability and the oxidation potential is further supported by the comparison of the transport properties of CNTFETs doped with the partially oxidized and partially reduced form of PANI (\textbf{E}) and pyridine. We found that CNTFETs (with 100~nm gate oxide) treated with PANI (\textbf{E}) has a V$_{th}$ 10~V more negative than the ones treated with pyridine. The difference in the V$_{th}$ shift correlates well with the different electron donating (oxidation potential) abilities of the two dopants.

In addition to utilizing different degrees of charge delocalization to modulate the charge transfer ability from dopants to CNTFETs, we also introduced electron-accepting groups to an n-doping molecule. For example, we combined the electron donating (amine) group and the electron accepting (phosphonic acid) group in one doping molecule (aminobuty- phosphonic acid (ABPA, H$_{2}$N(CH$_{2}$)$_{4}$PO(OH)$_{2}$)) and studied its impact on CNTFETs. In order to distinguish the effects of the two groups, we chose hexadecyl-phosphonic acid (HDPA, H$_{3}$C(CH$_{2}$)$_{15}$PO(OH)$_{2}$) as a control doping molecule where only the electron accepting phosphonic group is present. The pristine devices were fabricated as described above. They were then immersed in a 10~mM solution of ABPA and HDPA in ethanol, respectively. Fig.~3a shows the transfer characteristic before and after ABPA doping at V$_{ds}$ = $-$0.5~V. We observe a shift of the V$_{th}$ of $-$1.5~V, i.e. n-type doping. On the other hand, doping with HDPA in Fig.~3b shows that the device transfer characteristic shifted in the opposite direction and the device turned on at a more positive V$_{th}$, with a V$_{th}$ of $+$2.1~V. The above results show that the amine group plays a more important role in the charge transfer to CNTFETs than does the phosphonic acid group, which alone results in p-type doping. The p-doping effect of solid organic acids is stable in air, but very sensitive to moisture. Thus, all the measurements had been performed in a nitrogen atmosphere. A detailed analysis of acid p-doping of CNTFETs will be discussed in a future publication$^{24}$.

In conclusion, we demonstrated the use of the amine-containing monomers and polymers as dopants to convert p-type CNTFETs to n-type devices. We drastically improved device performance in both its ON- and OFF-states with hydrazine doping and attributed the improvement to metal-CNT contacts modification by the doping molecules. We also manipulated the charge donating ability by accessing different oxidation states of the amine-containing polymer polyaniline and demonstrated the importance of the oxidation state of amine-containing compounds to their charge donating ability. The readily oxidizable amine groups of PANI (\textbf{L}) are capable of transferring electrons to CNTFETs and therefore converting pristine p-type FETs to n-type. The more oxidized imine groups in PANI (\textbf{P}) behave as electron acceptor and convert the doped n-FETs back to p-type. 

The authors thank B. Ek and the CSS staff for expert technical assistance, N. Ruiz for assistance with e-beam lithography and M. Freitag and X. Qiu for insightful discussions. C. Klinke acknowledge gratefully the Swiss National Science Foundation (SNF) for their financial support. 

\section*{References}

(1) Avouris, Ph. MRS Bulletin 2004, 29, 403.

(2) McEuen, P. L.; Fuhrer, M. S.; Park, H. IEEE Trans. Nanotech. 2002, 1, 78.

(3) Javey, A.; Guo, J.; Farmer, D. B.; Wang, Q.; Wang, D.; Gordon, R. G.; Lundstrom, M.; Dai, H. Nano Lett. 2004, 4, 447. 

(4) Derycke, V.; Martel, R.; Appenzeller, J.; Avouris, Ph. Appl. Phys. Lett. 2002, 80, 2773.

(5) Kong J. and Dai, H., J. Chem. B 2001, 105, 2890-2893 

(6) Shim, M.; Javey, A.; Kam, N. W. S.; Dai, H. J. Am. Chem. Soc. 2001, 123, 11512.

(7) Bradley, K. ; Briman, M ; Star, A. ; and Gruner, G ; Nano Lett. 2004 4, 253-256.

(8) Star, A.; Han, T. R. ; Gabriel, J. C. P. ; Bradley, K.; Grüner, G. Nano Lett. 2003, 3, 1421. 

(9) Bradley, K.; Gabriel, J. C. P.; Briman, M.; Star, A.; Grüner, G. Phys. Rev. Lett. 2003, 91, 218301. 

(10) Radosavljevic, M.; Heinze, S.; Tersoff, J.; Avouris, Ph. Appl. Phys. Lett. 2003, 83, 2435.

(11) Collins, P. G.; Bradley, K.; Ishigami, M.; Zettl, A. Science 2000, 287, 1801.

(12) Cui, X. D.; Freitag, M.; Martel, R.; Brus, L.; Avouris, Ph. Nano Lett. 2003, 3, 783.

(13) Kong, J.; Chapline, M. G.; Dai, H. Adv. Mat. 2001, 13, 1384.

(14) Kim, W.; Javey, A.; Vermesh, O.; Wang, Q.; Li, Y.; Dai, H. Nano Lett. 2003, 3, 193.

(15) Peng, S.; Cho, K.; Qi, P.; Dai, H. Chem. Phys. Lett. 2004, 387, 271.

(16) Heinze, S.; Tersoff, J.; Martel, R.; Derycke, V.; Appenzeller, J.; Avouris, Ph. Phys. Rev. Lett. 2002, 89, 106 801.

(17) Takenobu, T.; Takano, T.; Shiraishi, M.; Murakami, Y.; Ata, M.; Kataura, H.; Achiba, Y.; Iwasa, Y. Nat. Mat. 2003, 2, 683.

(18) Nelsen, S.; Tran, H.; Ismagilov, R.; Ramm, M.; Chen, L. J. ; Powell, D. J. Org. Chem. 1998, 63, 2536.

(19) Radosavljevic, M.; Appenzeller, J.; Avouris, Ph.; Knoch, J. Appl. Phys. Lett. 2004, 84, 3693.

(20) Chen, J.; Klinke, C.; Afzali, A.; Avouris, Ph. IEDM, 2004.

(21) Among the three oxidation states of PANI, only the protonated salt form of emeraldine is conductive, which is not used in this work. The other forms are insulating and will not form leakage path in-between the devices.

(22) Afzali, A.; Buchwalter, S. L.; Buchwalter, L. P.; Hougham, G. Polymer 1997, 38, 4439. 

(23) The OFF currents of all devices are around sub-pA range.

(24) Klinke, C.; Chen, J.; Afzali, A.; Avouris, Ph. submitted.

\end{document}